\documentclass[journal]{IEEEtran}
\usepackage{graphicx}
\usepackage{subfigure}
\usepackage{amsmath}
\usepackage{theorem}
\usepackage{cite}
\usepackage{color}

\newtheorem{lemma}{\textbf{Lemma}}

\newtheorem{DD}{Definition}
\usepackage[]{algorithm2e}
\begin{document}
\title{Joint Rate Selection and Wireless Network Coding for Time Critical Applications}
\author{
\small Xiumin Wang$^\star$, Chau Yuen$^\star$, Yinlong Xu$^\dagger$\\
$^\star$ Singapore University of Technology and Design, Singapore\\
$^\dagger$ School of Computer Science, University of Science and Technology of China, China\\
\small Email: wangxiumin@sutd.edu.sg, yuenchau@sutd.edu.sg, ylxu@ustc.edu.cn
\thanks{This research is supported by the International Design Center at Singapore University of Technology and Design, Singapore (Grant No. IDG31100102 \& IDD11100101).}}
\maketitle

\begin{abstract}
In this paper, we  dynamically select the transmission rate and design wireless network coding to improve the quality of services such as delay for time critical applications.
With low transmission rate, and hence longer transmission range, more packets may be encoded together, which increases the coding opportunity. However, low transmission rate may incur extra transmission delay, which is intolerable for time critical applications.
We design a novel joint rate selection and wireless network coding (RSNC) scheme with delay constraint, so as to minimize the total number of packets that miss their deadlines at the destination nodes. We prove that the proposed problem is NP-hard, and propose a novel graph model and transmission metric which consider
both the heterogenous transmission rates and the packet deadline constraints during the graph construction. Using the graph model, we mathematically formulate the problem and design an efficient
algorithm to determine the transmission rate and coding strategy for each transmission. Finally, simulation results demonstrate the superiority of the RSNC scheme.
\end{abstract}

\vspace{-0.1in}
\section{Introduction}
With the increase in both wireless channel bandwidth and the computational capability of wireless devices, wireless networks
now can be used to support time critical applications such as video streaming or interactive gaming. Such time critical applications require the data content to reach the destination node(s) in a timely fashion, i.e., a delay deadline is imposed on packet reception, beyond which the reception becomes useless (or invalid) \cite{XTL2006Time-critical14}.


Recently, network coding becomes a promising approach to improve wireless network performance \cite{Sagduyu2007,ACL+2000Network1216,KRH+2008XORs510}. Specifically, the work in \cite{KRH+2008XORs510} proposed the first network coding based packet forwarding architecture, named {\em COPE}, to improve the throughput of wireless networks. With COPE, each node opportunistically overhears some of the packets transmitted by its neighbors, which are not intended to itself. The relay node can then intelligently XOR multiple packets and forward it to multiple next hops with only one transmission, which results in a significant throughput improvement.

In most recent works for wireless network coding, network nodes always transmit packets at a fixed rate. However, most wireless systems are now capable of performing adaptive modulation to vary the link transmission rate in response to the
signal to interference plus noise at the receivers. Transmission rate diversity exhibits a rate-range tradeoff: the higher the transmission rate, the shorter the transmission range for a given transmission power \cite{KV2009Is646}. To aid overhearing, one may use the lowest transmission rate, so as to successfully deliver packet to more receivers/overhearing nodes.
Although this may increase the coding opportunity, it may not yield good performance, especially for time critical applications, as the arrival times of packets may be delayed.

In the literature, only a few works studied the relationships between adapting the transmission rate and the network coding gain \cite{KV2009Is646,CJH2010Joint2444,Kim2010,Ni2008}. The work in \cite{KV2009Is646} showed that compared with pure network coding scheme, joint rate adaptation and network coding is more effective in throughput performance. They also proposed a joint rate selection and coding scheme to minimize the sum of the uplink and the downlink costs in star network topology. The work in \cite{CJH2010Joint2444} mathematically formulated the optimal packet coding and rate selection problem as an integer programming problem, and proposed an efficient heuristic algorithm to jointly find a good combination of coding solution and the transmission rate. However, there are only a few works considered the delay guarantee of packet receptions, which is especially important for time critical applications. So far, only \cite{ZX2010Broadcast6} considered the delay constraint of packet reception, and proposed a coding scheme to minimize the number of packets that miss their deadlines. However, they assume that the transmission rates on all the links are fixed and the same.

In this paper, by considering the impact of both transmission rate and network coding on the packet reception delay, we design a joint rate selection and network coding (RSNC) scheme for wireless time critical applications, so as to minimize the total number of packets that will miss their deadlines at the destination nodes. The main contributions of our paper can be concluded as follows.
\begin{itemize}
\item We propose a novel graph model, which considers both the heterogenous transmission rates and the deadlines of the packet receptions during the graph construction. Based on the graph model, we mathematically formulate the problem of minimizing the total number of packets that miss their deadlines by joint rate selection and network coding, as an integer programming problem.
\item We propose a metric to determine the coded packet and the transmission rate for each packet transmission. By considering the impact of the transmission rate on both delay and network coding gain, we also design an efficient algorithm to optimize the proposed metric.


\item We compare the performance of the proposed RSNC scheme with some existing algorithms. Simulation results show that the proposed scheme can significantly reduce the packet deadline miss ratio.
\end{itemize}

The rest of the paper is organized as follows. We define our problem in Sec.~\ref{Sec.formulation}. Sec.~\ref{Sec.solution} gives the graph model and problem formulation. The algorithm design for each transmission is given in Sec.~\ref{Sec.algorithm}. We show the simulation results in Sec.~\ref{Sec.simulation}, and conclude the paper in Sec.~\ref{Sec.conclusion}.

\begin{figure}[t]
\begin{center}\vspace{-0.04in}
\includegraphics[height=27mm,width=80mm]{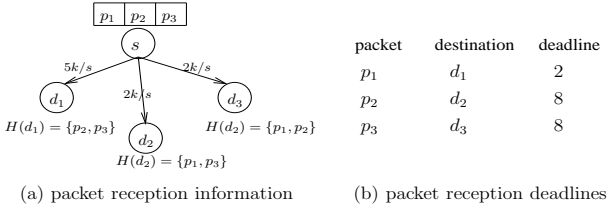}\vspace{-0.1in}
\caption{Motivation illustration}\vspace{-0.08in} \label{Fig.rsnc}
\end{center}\vspace{-0.15in}
\end{figure}

\vspace{-0.1in}
\section{Problem Definition}\label{Sec.formulation}
In this section, we first illustrate the motivation of our problem. We then give the problem description and its complexity.
\vspace{-0.24in}
\subsection{Motivation Illustration}\label{Sec.formulation.motivation}\vspace{-0.02in}
We now give an example to show how joint rate selection and network coding affect the time critical applications.

Take Fig.~\ref{Fig.rsnc} as an example, where source/transmitting node node $s$ needs to transmit packet $p_1,p_2,p_3$ to node $d_1,d_2,d_3$
respectively. Fig.~\ref{Fig.rsnc}(a) gives the set of overheard packets $H(d_i)$ at destination $d_i\in \{d_1,d_2,d_3\}$. Suppose that the size of each packet is $B=10k$, and the maximum transmission rates from $s$ to $d_1,d_2,d_3$ are $5k/s,2k/s$ and $2k/s$, respectively.
Fig.~\ref{Fig.rsnc}(b) shows the reception deadline of each ``wanted" packet at its destination. For the current transmission, according to the work in \cite{KRH+2008XORs510,ZX2010Broadcast6}, $s$ will send the encoded packet $p_1\oplus p_2\oplus p_3$, as the most number of destinations can decode it. However, there is a problem for selecting the transmission rate at $s$. If $5k/s$ is selected, $d_2,d_3$ can not successfully receive the packet, as the maximum transmission rates from $s$ to them are both $2k/s$. If $2k/s$ is selected, although all of the three receivers $d_1,d_2,d_3$ can receive and decode one ``wanted" packet, $p_1$  will miss its deadline at $d_1$, as its arrival time is $\frac{10k}{2k/s}=5s$.

As an alternative, we may choose to first send packet $p_1$ with transmission rate $5k/s$, where destinations $d_1$ will obtain a ``wanted" packet in $2s$. After this transmission, the encoded packet $p_2\oplus p_3$ can be sent with transmission rate $2k/s$, where destination $d_2$ and $d_3$ will obtain a ``wanted" packet after $7s$ (including the waiting time of the first transmission). Obviously, the latter solution is better than the first one, as no packet will miss their deadline.

\vspace{-0.15in}
\subsection{Problem Description}\vspace{-0.02in}
In this paper, we consider the application of network coding in wireless networks. Each network node knows the overheard/routed packets that its neighbors have such that it can perform network coding operations. Such information can be achieved by using {\em reception reports}, as introduced in \cite{KRH+2008XORs510}. We also assume that the forwarding/relaying node knows the deadlines of the packet receptions at its receivers. Specifically, we consider the transmission scheme within a single hop since multi hop can be regarded as multiple single hops. As in COPE \cite{KRH+2008XORs510}, only XORs coding is performed at the node in our work.

Without loss of generality, let $s$ be the current source node, and $P=\{p_1,p_2,\cdots,p_n\}$ be the set of packets required to be transmitted from $s$. Suppose that $D=\{d_1,d_2,\cdots,d_m\}$ is the set of $s$'s neighbors which requires packets in $P$ sent from node $s$. Let $R(d_i)$ be the set of ``wanted" packets at $d_i$, and $H(d_i)$ be the set of overheard packets at $d_i$, where $R(d_i)\subseteq P,H(d_i)\subseteq P$. For each $p_j\in R(d_i)$, let $T(d_i,p_j)$ be the reception deadline of packet $p_j$ at node $d_i$. We also assume that $r(s,d_i)$ is the maximum transmission rate on link $(s,d_i)$, and only if the transmission rate from $s$ to $d_i$ is less than $r(s,d_i)$, the packet sent from $s$ can be successfully received by $d_i$ \cite{KV2009Is646}. We also assume that the size of each packet is $B$.

Our problem is that given the set of overheard packets at each node $d_i$, $H(d_i)$, the set of packets required by $d_i$, $R(d_i)$, the deadline of required packet $p_j$ at node $d_i$, $T(d_i,p_j)$, and the maximum transmission rate $r(s,d_i)$ on the link from $s$ to $d_i$, design the encoding strategy of the packets and select the transmission rate for each propagation, such that the total number of packets that miss their deadlines at each destination is minimized.

Let $z_{i,j}$ be $1$ if packet $p_j$ misses its deadline at $d_i$, otherwise, let it be $0$, where $p_j\in R(d_i)$. Thus, our objective is to minimize
$$\sum_{d_i\in D}\sum_{p_j\in R(d_i)}z_{i,j}$$ In this paper, we refer such a problem of joint Rate Selection and Network Coding (RSNC) for time critical applications as RSNC problem.

\vspace{-0.1in}
\subsection{Problem Complexity}
\begin{lemma}
The RSNC problem is NP-hard.
\end{lemma}\vspace{-0.1in}
\begin{proof}
We can consider a special case of the RSNC problem: $T(d_i,p_j)=1$ and the maximum transmission rates on all the links are the same. Then, this special case is equivalent to finding a maximum weight clique problem as in \cite{ZX2010Broadcast6}, which is known as an NP-hard problem. Thus, the RSNC problem is also NP-hard.
\end{proof}

\vspace{-0.1in}
\section{Graph Model and RSNC Formulation}\label{Sec.solution}
\subsection{Graph Model}\label{graph.model}
Although the graph model in \cite{ZX2010Broadcast6} works well for the case where the transmission rates on all the links are the same and fixed, it can not be used directly for our RSNC problem. Here, we construct a novel graph model $G(V,E)$, which considers both the transmission rates and the packet reception deadlines.

We define $r_{min}(s,d_i|p_j)=\frac{B}{T(d_i,p_j)}$ as the minimum transmission rate that can be used to meet the deadline of $p_j\in R(d_i)$ at $d_i$.
We add a vertex $v_{i,j}$ in $V(G)$, only if the following two conditions can be met.

(1) $p_j\in R(d_i)$;

(2) $r_{min}(s,d_i|p_j)\leq r(s,d_i)$.

Note that, if $r_{min}(s,d_i|p_j)> r(s,d_i)$, packet $p_j$ will definitely miss its deadline at $d_i$.
Thus, conditions (1) and (2) ensure that we add a vertex $v_{i,j}$ in $V(G)$ only if the ``wanted" packet $p_j$ will not miss its deadline at $d_j$.
That is, $V(G)=\{v_{i,j}|p_j\in R(d_i), r_{min}(s,d_i|p_j)\leq r(s,d_i)\}$.

Then, for any two different vertices $v_{i,j},v_{i',j'}\in V(G)$, there is a link $(v_{i,j},v_{i',j'})\in E(G)$ if all the following conditions can be satisfied.

(a) $i\neq i'$;

(b) $j=j'$ or $p_j\in H(d_{i'})$ and $p_{j'}\in H(d_i)$;

(c) $r_{min}(s,d_i|p_j)\leq r(s,d_{i'})$ and $r_{min}(s,d_{i'}|p_{j'})\leq r(s,d_i)$.

For any clique $Q=\{v_{i_1,j_1},v_{i_2,j_2},\cdots\}$ in $G$, let $P'=\{p_j|v_{i,j}\in Q\},D'=\{d_i|v_{i,j}\in Q\}$. According to the work in \cite{ZX2010Broadcast6}, if node $d_i\in D'$ successfully receives the encode packet $p_{j_1}\oplus p_{j_2}\oplus \cdots\oplus p_{|P'|}$, where $p_{j_1}, p_{j_2},\cdots,p_{|P'|} \in P'$, $d_i$ can decode a ``wanted" packet $p_j$, where $v_{i,j}\in Q$.

Next, we will use an example to show the novelty of our graph model as compared to others in the literature, e.g., \cite{ZX2010Broadcast6}.

\begin{figure}[t]
\begin{center}
\includegraphics[height=23mm,width=70mm]{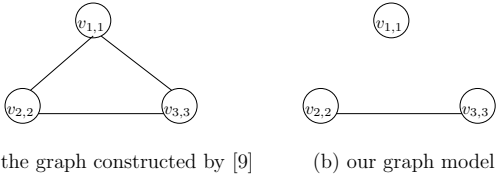}\vspace{-0.08in}
\caption{Different graph model comparison}\vspace{-0.15in}\label{Fig.example}
\end{center}\vspace{-0.1in}
\end{figure}

Still take Fig.~\ref{Fig.rsnc} as an example. The graph constructed by \cite{ZX2010Broadcast6} is shown in Fig.~\ref{Fig.example} (a). According to \cite{ZX2010Broadcast6}, any clique in the graph represents a feasible encoded packet. Thus, $p_1\oplus p_2\oplus p_3$ can be sent and its intended next hops are $d_1,d_2,d_3$, because $\{v_{1,1}$,$v_{2,2}$,$v_{3,3}\}$ forms a clique. As described in Sec.~\ref{Sec.formulation.motivation}, it is not a good choice.
However, with our graph model shown in Fig.~\ref{Fig.example}(b), $p_1,p_2, p_3$ will not be encoded as vertices
$v_{1,1}$,$v_{2,2}$,$v_{3,3}$ do not form a clique in the graph. In addition, for the current transmission, the encoded packet derived from any clique in the graph can be sent without missing the deadlines at its intended destinations. For example, if $p_2\oplus p_3$, which is derived from the clique $\{v_{2,2},v_{3,3}\}$, is sent with the minimum of the maximum transmission rates among $r(s,d_2)$ and $r(s,d_3)$, $2k/s$, its intended next hops $d_2,d_3$ can successfully decode the packets $p_2,p_3$ respectively, without missing their deadlines.

With the graph $G(V,E)$, we have the following lemma.
\vspace{-0.02in}\begin{lemma}\vspace{-0.1in}\label{lemma_graph}
For the current packet transmission, if the encode packet $p_{j_1}\oplus p_{j_2}\oplus \cdots\oplus p_{|P'|}$, where $p_{j_1},\cdots,p_{|P'|} \in P'$, is sent with the transmission rate $r=\min\{r(s,d_i)|d_i\in D'\}$, it will be received by all the nodes in $D'$. In addition, for each $v_{i,j}\in Q$, the packet $p_j$ will be decoded by $d_i$ without missing its deadline.
\end{lemma}\vspace{-0.02in}
\begin{proof}
Firstly, we can easily obtain that with transmission rate $r=\min\{r(s,d_i)|d_i\in D'\}$, all the receivers in $D'$ can successfully receive the sending packet. This is because the transmission rate $r$ must be lower than the maximum transmission rate from $s$ to $d_i\in D'$.

Secondly, our graph is the subgraph of that constructed in~\cite{ZX2010Broadcast6}. According to \cite{ZX2010Broadcast6}, if $p_{j_1}\oplus p_{j_1}\oplus \cdots\oplus p_{j_{|P'|}}$ is successfully received by $d_i \in D'$, $d_i$ can decode its ``wanted" packet $p_j$, where $v_{i,j}\in Q$.
Thus, any receiver $d_i\in D'$ can obtain a ``wanted" packet $p_j$ from $p_{j_1}\oplus p_{j_2}\oplus \cdots\oplus p_{j_{|P'|}}$ with transmission rate $r$, where $v_{i,j}\in Q$.

Thirdly, according to the condition (c), we have
{\small\begin{eqnarray}
r_{min}(s,d_i|p_j)\leq \min_{d_i\in D'}\{r(s,d_i)\}=r
\end{eqnarray}}
So, its arrival time at receiver $d_i$ is
{\small\begin{eqnarray}
\frac{B}{r}\leq \frac{B}{r_{min}(s,d_i|p_j)}=\frac{B}{\frac{B}{T(d_i,p_j)}}=T(d_i,p_j)
\end{eqnarray}}
In other words, the arrival time of the packet $p_j\in P'$ will not miss its deadline at its receiver $d_i\in D'$, where $v_{i,j}\in Q$.
\end{proof}

As in Lemma~\ref{lemma_graph}, a clique $Q$ in the graph represents a feasible transmission solution for the current propagation, with the encoded packet $p_{j_1}\oplus p_{j_2}\oplus \cdots\oplus p_{j_{|P'|}}$, transmission rate $r=\min\{r(s,d_i)|d_i\in D'\}$, intended next hops in $D'$, and the propagation delay $\frac{B}{r}$.

\vspace{-0.1in}
\subsection{RSNC Formulation}
While Lemma~\ref{lemma_graph} ensures that any encoding strategy based on any clique in the graph will be delivered within deadline for the current packet transmission, the transmission orders of the encoded packets, represented by the cliques in $G(V,E)$, also affects the timely packet receptions at their destinations.

As shown in Fig.~\ref{Fig.example}(b), if we first schedule packet $p_1$ with transmission rate $5k/s$, represented by clique $\{v_{1,1}\}$, and then schedule packet $p_2\oplus p_3$ with transmission rate $2k/s$, represented by clique $\{v_{2,2},v_{3,3}\}$, all the packets will be received at their destinations without missing deadlines. However, if we first schedule packet $p_2\oplus p_3$, and then packet $p_1$, packet $p_1$ will miss its deadline at $d_1$.

Thus, the next task for us is to find a set of cliques in the graph and schedule the transmissions of the encoded packets represented by each clique, so as to minimize the number of missed packets for the whole transmission process.

Suppose that $Q_h=\{v_{i_1,j_1},v_{i_2,j_2},\cdots,\}$ is a clique in the graph, and the encoded packet represented by it is sent as the $h$-th transmission at node $s$. We also assume that $P'_h=\{p_j|v_{i,j}\in Q_h\}$, $D'_h=\{d_i|v_{i,j}\in Q_h\}$. Thus, the $h$-th transmission at $s$ is $p_{j_1}\oplus p_{j_2}\oplus\cdots \oplus p_{j_{|P'_h|}}$ where $p_{j_1},p_{j_2},\cdots,p_{j_{|P'_h|}}\in P'_h$, and the transmission rate is $r_h=\min_{d_i\in D'_h}\{r(s,d_i)\}$. Let $T_h$ be the transmission delay of the $h$-th transmission.

We firstly define the following variant.
{\small \begin{equation}
x_{i,j,h}=\left\{
\begin{aligned}
&1,\mbox{if vertex}~v_{i,j}~\mbox{is included in clique}~Q_h\\
&0,\mbox{otherwise}
\end{aligned}
\right.\\
,\forall v_{i,j}\in V(G)
\end{equation}}
Then, we can formulate the RSNC problem based on the graph model as follows.
{\small \begin{eqnarray}\label{formulation.obj}
\min_{Q_h} & &\sum_{d_i\in D}\sum_{p_j\in R(d_i)}z_{i,j}
\end{eqnarray}}
subject to{\small \begin{eqnarray}
&&\sum_{h=1}^{|V(G)|}x_{i,j,h}=1,\forall v_{i,j}\in E(G)\label{ch.1}\\
&&x_{i,j,h}+x_{i',j',h}=1,\forall (v_{i,j},v_{i',j'})\notin E(G)\label{ch.2}\\
&&T_h=\max_{v_{i,j}\in V}\{\frac{B*x_{i,j,h}}{r(s,d_i)}\},1\leq h\leq |V(G)|\label{ch.3}\\
&&\sum_{h=1}^{|V(G)|}(x_{i,j,h}*\sum_{j=1}^hT_h)\leq T(d_i,p_j)+\xi z_{i,j},\forall v_{i,j}\label{ch.4}\\
&&\sum_{h=1}^{|V(G)|}(x_{i,j,h}*\sum_{j=1}^hT_h)\geq T(d_i,p_j)-\xi (1-z_{i,j}),\forall v_{i,j}\label{ch.5}
\end{eqnarray}}\vspace{-0.04in}
where $\xi$ is a sufficient large constant.

In the above formulation, the term of the objective represents the number of packets that miss their deadlines at the receivers, which needs to be minimized.
Constraint~(\ref{ch.1}) denotes that each vertex in the graph can only belong to one clique. Constraint~(\ref{ch.2}) means that if there is no link between vertex $v_{i,j}$ and $v_{i',j'}$, vertices $v_{i,j},v_{i',j'}$ can not be in the same clique. Constraint~(\ref{ch.3}) gives the transmission delay for the $h$-th transmission, which is equal to the transmission delay with the minimum transmission rate among the rates from $s$ to all intended receivers. The sufficient large constant $\xi$ is used to guarantee that if $\sum_{h=1}^{|V|}(x_{i,j,h}*\sum_{j=1}^hT_h)>T(d_i,p_j)$, $z_{i,j}$ must be $1$, as denoted in Constraint~\ref{ch.4}, and if $\sum_{h=1}^{|V|}(x_{i,j,h}*\sum_{j=1}^hT_h)\leq T(d_i,p_j)$, $z_{i,j}$ must be $0$, as denoted in Constraint~\ref{ch.5}. Note that the arrival time of the packet in the $h$-th transmission should consist of both the waiting time of the previous $h-1$ transmissions and the transmission time of the $h$-th transmission, i.e., $\sum_{j=1}^h T_h$.
Thus, Constraint~(\ref{ch.4}) and~(\ref{ch.5}) show that only if the arrival time of $p_j$ at $d_i$, i.e., $\sum_{h=1}^{|V|}(x_{i,j,h}*\sum_{j=1}^hT_h)$, is no more than the reception deadline of $p_j$ at $d_i$, $z_{i,j}$ can be $0$.

With the above integer nonlinear programming, we can get the optimal solution of RSNC problem. However, the computational complexity for the above integer nonlinear programming is too high when the graph is large. Thus, we need to find an efficient algorithm to solve it.

\vspace{-0.04in}
\section{Joint Rate Selection and Network Coding Algorithm}\label{Sec.algorithm}\vspace{-0.03in}
Since each clique in the graph represents a feasible transmission strategy for the current transmission, instead of determining the whole transmission sequence at once, we first design the algorithm to determine the encoding strategy and rate selection scheme for each packet propagation, by selecting a clique at a time. The whole transmission process consists of multiple packets transmission/cliques selection.
\vspace{-0.07in}
\subsection{Metric Consideration for Each Packet Propagation}\vspace{-0.03in}
First of all, in order to measure the ``goodness" of transmitting an encoded packet at a specific transmission rate for each packet propagation, it is necessary for us to adopt a reasonable metric which should take into account the impact of the transmission rate and the packet reception deadlines.
In this section, we shall design a metric, which not only satisfies as more requests as possible, but also minimizes the number of packets missing the deadlines after the current transmission.

For the current transmission, given an encoded packet and a selected transmission rate, let $f_{i,j}$ be $1$ if $p_j$ is decoded/received by $d_i$ from the current propagation without missing its deadline, otherwise, let it be $0$. In addition, as described in Fig.~\ref{Fig.rsnc}, the current encoding strategy and transmission rate also affect the timely receptions of the packets in the following propagations. Let $l_{i,j}$ be 1 if $p_j$ will definitely miss its deadline at $d_i$ after the current propagation, otherwise, let it be $0$. Later, we will introduce how to calculate $f_{i,j}$ and $l_{i,j}$ for a given encoded packet and transmission rate. Let $r$ be the transmission rate selected for the current propagation.

Our metric can be defined as follows.
\begin{DD}
For a coding solution $p_{j_1}\oplus p_{j_2}\oplus \cdots \oplus p_{j_L}$, define the metric $U$ when using the transmission rate $r$ as follows:
\vspace{-0.05in}
{\small\begin{eqnarray}\label{equation.U}
U=\sum_{d_i\in D}\sum_{p_j\in R(d_i)} \alpha_j f_{i,j}- \sum_{d_i\in D}\sum_{p_j\in R(d_i)}\alpha_j l_{i,j}
\end{eqnarray}}
where $\alpha_j$ is the parameter, which can be defined as the benefit (e.g., importance) of the packet $p_j$.
\end{DD}\vspace{-0.05in}

Firstly, for a given encoded packet $p_{j_1}\oplus p_{j_2}\oplus \cdots \oplus p_{j_L}$, $f_{i,j}$ is $1$ only if all the following conditions are met: 1) $r\geq r(s,d_i)$, which means $d_i$ can successfully receive the sending packet; 2) $p_j\in R(d_i)$, which means $p_j$ is required at $d_i$; 3) All the other packets combined in the encoded packet except $p_j$ are available at $d_i$, which denotes the decoding requirement of $p_j$ at $d_i$; 4) $\frac{B}{r}\leq T(d_i,p_j)$, which shows the requirement of the reception deadline.
Secondly, for each packet $p_j\in R(d_i)$ which is not successfully received/decoded by $d_i$ from the current transmission without missing deadlines ($f_{i,j}=0$), $l_{i,j}$ is $1$ only if
{\small \begin{eqnarray} \label{eq.miss}
\frac{B}{r}+\frac{B}{r(s,d_i)}>T(d_i,p_j)
\end{eqnarray}}
Here, $\frac{B}{r}$ is the transmission delay of the current transmission, and $\frac{B}{r(s,d_i)}$ denotes the minimum delay to meet $p_j$'s deadline at $d_i$ in the next transmission. If the sum of the current transmission delay and the next minimum transmission delay is larger than the deadline of $p_j$ at $d_i$, $p_j$ will definitely miss its deadline, i.e., $l_{i,j}=1$.
Thus, given an encoded packet and its transmission rate, $f_{i,j}$ and $l_{i,j}$ are both determined.

Hence the meaning of metric $U$ in (\ref{equation.U}) can be explained as follows.
The first term $\sum_{d_i\in D}\sum_{p_j\in R(d_i)}\alpha_j f_{i,j}$ denotes the benefit obtained from the packets that are received without missing their deadlines from the current transmission. The second term $\sum_{d_i\in D}\sum_{p_j\in R(d_i)}\alpha_j l_{i,j}$ represents the lost due to packets that will definitely miss their deadlines after the current transmission. So, the metric $U$ denotes the net benefit obtained from the current encoded packet and the transmission rate.
Thus, for each packet propagation, we aim to determine an encoded packet and select the transmission rate $r$, that will maximize the metric $U$.

Note that, the problem of maximizing the defined metric $U$ is also NP-hard. We can prove it by considering its special case: the transmission rates on all the links are the same,  the reception deadlines for all the packets are the transmission time of one packet, and each packet has the same benefit $\alpha_j$. The special case of maximizing the defined metric $U$ becomes to maximize the total number of the receivers that can decode one ``wanted" packet from the current encoded packet, which has been proved to be NP-hard in \cite{XiuminWang2010}.

\vspace{-0.1in}
\subsection{Heuristic Algorithm Design for Each Packet Propagation}\label{Sec.algorithm.design}
Although maximizing the defined metric $U$ is also NP-hard, we can easily obtain the following observations, based on which we can design the heuristic algorithm.

P1: Maximizing the first term of the metric $U$ is equal to find a maximum weight clique in the graph, where the weight at vertex $v_{i,j}$ is defined as the benefit $\alpha_j$.

P2: The transmission rate is a parameter that adjusts the trade-off between delay and network coding gain. If $s$ uses a low transmission rate, more receivers can successfully receive the sending packet, and the current transmission may satisfy more receivers' requirements, denoted by the first term in $U$. However, low transmission rate means high transmission delay, which may cause more packets to miss their deadlines in the following transmission, denoted by the second term in $U$.

Based on the above observations, we then design a heuristic algorithm for each packet propagation, by gradually increasing the transmission rate. Initially, the transmission rate is set to be no less than the lowest one from $s$ to its receivers. Let $TR=\{r(s,d_i)|d_i\in D\}$ be the set of available transmission rates from $s$ to all the destinations, and let $Tr_k$ be the $k$-th lowest rate in $TR$. As in Sec.~\ref{graph.model}, we construct the auxiliary graph with the given information. We also assign the weight $\alpha_{j}$ in vertex $v_{i,j}$ for $\forall i$ to denote the benefit of $p_j$.

In the $k$-th step, we restrict that the transmission rate used at $s$ must be no less than $Tr_k$. For $d_i$, if its maximum transmission rate from $s$ is less than $Tr_k$, it can not successfully receive the sending packet. This restriction can be realized by omitting the vertex $v_{i,j}$ in $G(V,E)$ if $r(s,d_i)< Tr_k$. Then, we find the maximum weight clique in the subgraph $\{v_{i,j}|r(s,d_i)\geq Tr_k,v_{i,j}\in V(G)\}$, and adopt the transmission rate represented by the found clique. Each vertex $v_{i,j}$ in the found clique denotes that $p_j$ will be successfully obtained by $d_i$ without missing its deadline, for the given transmission rate. For each of the other packets that can not be obtained at their receivers from the current transmission, we then judge whether it will definitely miss its deadline at its destinations, by (\ref{eq.miss}). Thus, in each step, we calculate $U$. Such process continues until all the rates in $TR$ are considered. Finally, we compare the values of $U$ obtained from each step and adopt the one with the largest value as solution.
Note that, if there are more than one solution with the maximum value of $U$, we will choose the one with the smaller lost represented by the second term of (\ref{equation.U}).
The detailed of the algorithm is shown in Algorithm 1 of Fig.~\ref{alg}.
\vspace{-0.1in}
\subsection{Algorithm for the Whole Transmission Process}\vspace{-0.02in}
While algorithm 1 in Sec.~\ref{Sec.algorithm.design} describes the encoding of packets and selection of rate for every transmission, the whole transmission process will consist of multiple of such single process. We will first construct the graph $G(V,E)$ based on model in Sec.~\ref{graph.model}, and the graph will be updated by removing the selected vertices in the found clique by Algorithm 1, and the vertex $v_{i,j}$ if $p_j$ will definitely miss its deadline at destination $d_i$. The packet reception deadlines for the packets also need to be updated after each transmission. The whole transmission process continues until the vertices set $V$ of $G$ becomes empty. The detail algorithm for the whole transmission is given in Algorithm 2 of Fig.~\ref{alg}.

\begin{figure}[t]\center
\scriptsize{
\begin{tabular}{|l|}
\hline {\bf Algorithm 1: one packet propagation process}\\
\hspace{2mm} $U_k=0$, $\forall k\in \{1,2,\cdots,|TR|\}$;\\
\hspace{2mm} $f^k_{i,j}=l^k_{i,j}=0$, $\forall k\in \{1,2,\cdots,|TR|\},v_{i,j}\in V(G)$;\\
\hspace{2mm}{\bf for} $k\longleftarrow 1$ to $|TR|$\\
\hspace{4mm} find max weight clique $Q$ in subgraph $\{v_{i,j}|r(s,d_i)\geq Tr_k, v_{i,j}\in V\}$;\\
\hspace{4mm} $f^k_{i,j}=1$, if $v_{i,j}\in Q$, for $\forall i,j$;\\
\hspace{4mm} $r'_k=\min_{v_{i,j}\in Q}\{r(s,d_i)\}$;\\
\hspace{4mm} {\bf For} each $v_{i,j}\in V(G),v_{i,j}\notin Q$\\
\hspace{6mm} $l^k_{i,j}=1$, if $\frac{B}{r'_k}+\frac{B}{r(s,d_i)}>T(d_i,p_j)$;\\
\hspace{4mm} {\bf Endfor}\\
\hspace{4mm} $U_k=\sum_{d_i\in D}\sum_{p_j\in R(d_i)} \alpha_j f^k_{i,j}- \sum_{d_i\in D}\sum_{p_j\in R(d_i)}\alpha_j l^k_{i,j}$;\\
\hspace{2mm}{\bf Endfor}\\
\hspace{2mm} add $k$ into $W$ if $U_k$ is the maximum among $\{U_k|0\leq k\leq |TR|\}$;\\
\hspace{2mm} $k'=\arg\min_{k\in W}\{\sum_{d_i\in D}\sum_{p_j\in R(d_i)}\alpha_j l^k_{i,j}\}$;\\
\hspace{2mm} $U=U_{k'}$;$f_{i,j}=f^{k'}_{i,j};l_{i,j}=l^{k'}_{i,j}$;\\
\hspace{2mm} the encoded packet is $\bigoplus_{v_{i,j}\in Q}p_j$ for the current transmission;\\
\hspace{2mm} $r=\min_{v_{i,j}\in Q}\{r(s,d_i)\}$;\\
\hline
\hline {\bf Algorithm 2: the whole packet transmission process}\\
\hspace{2mm} construct graph $G(V,E)$;\\
\hspace{2mm} {\bf while} ($V(G)$ is not empty)\\
\hspace{4mm} conduct Algorithm 1 for the current packet propagation;\\
\hspace{4mm} remove the selected clique from $G(V,E)$;\\
\hspace{4mm} remove the vertex $v_{i,j}$ from $V(G)$ if $l_{i,j}=1$;\\
\hspace{4mm} update the packet reception deadline, e.g., $T(d_i,p_j)=T(d_i,p_j)-\frac{B}{r}$;\\
\hspace{4mm} update $G(V,E)$ based on the remaining $V(G)$ and $E(G)$;\\
\hspace{2mm}{\bf Endwhile}\\
\hline
\end{tabular}\vspace{-0.05in}
\caption{Algorithm Design}\vspace{-0.15in} \label{alg}}
\end{figure}

\vspace{-0.1in}
\section{Simulation Results}\label{Sec.simulation}\vspace{-0.02in}
In this section, we demonstrate the effectiveness of our RSNC scheme through simulations.
We randomly generate a set of available packets in $H(d_i)$ and the ``wanted" packets in $R(d_i)$ at destination $d_i\in D$, where $H(d_i)\bigcap R(d_i)=\emptyset$. The maximum transmission rate from $s$ to $d_i$ is randomly selected in $[rmin,rmax]$, and the packet reception deadline is randomly generated in $[Tmin,Tmax]$.

For comparison purpose, we include two baseline algorithms,
namely, {\em DSF (deadline smallest first) coding} algorithm \cite{ZX2010Broadcast6} and {\em SIN-1} algorithm \cite{XTL2006Time-critical14}. DSF coding algorithm does not consider the heterogenous transmission rates on the links, and in each time slot, always finds the maximum weight clique in the defined graph. SIN-1 algorithm always sends the packet with the minimum ``SIN-1" in each transmission, where ``SIN-1" of packet $p_j$ is defined as the ratio of the duration from the current time to the deadline of the most urgent request for packet $p_j$, to the number of requests for $p_j$. In the simulation, we compare the deadline miss ratio under different transmission schemes, where deadline miss ratio is defined as the ratio of the number of packets missing their deadlines to the total number of requests.
For each setting, we present the average result of 100 samples.

\vspace{-0.1in}
\subsection{The Impact of the Transmission Rate}\vspace{-0.02in}
We first investigate the impact of the transmission rate on the performance of random one packet propagation during the whole transmission process. Given the rate for the current transmission, we run the maximum weight clique algorithm in the graph model to find the maximum number of packets that can be obtained at their destinations without missing deadline, i.e., satisfied requests, based on which we derive the number of packets that will definitely lose their deadlines in the next transmissions according to (\ref{eq.miss}), i.e., failed requests. We set $\alpha_j=1,n=10,m=20$, $rmin=10,rmax=100$, and $Tmin=10,Tmax=50$.
\begin{figure}[h]
\begin{center}\vspace{-0.1in}
\includegraphics[height=35mm,width=63mm]{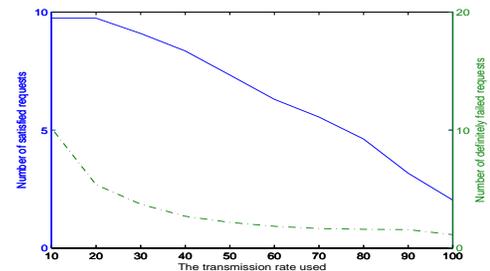}\vspace{-0.1in}
\caption{The impact of the transmission rate on the performance of one transmission.}\vspace{-0.15in} \label{sim.tradeoff}
\end{center}\vspace{-0.1in}
\end{figure}

As shown in Fig.~\ref{sim.tradeoff}, with the increase of the transmission rate, the number of packets that can be successfully received/decoded by their receivers from the current transmission decreases. The reason is that with the increase of the transmission rate, higher number of destinations may not receive the packets due to shorter transmission range, which thus decreases the encoding opportunity at $s$. We can also see that with the increase of the transmission rate, the number of packets that will definitely miss their deadlines at the receivers decreases. This is because, higher transmission rates incur less transmission delay, which gives more chances for the other packets received timely in the following transmissions. The above observation motivates the algorithm design in Sec.~\ref{Sec.algorithm.design}, by deciding the tradeoff between transmission rate and network encoding strategy.

\begin{figure}[t]
\begin{center}\vspace{-0.07in}
\includegraphics[height=35mm,width=60mm]{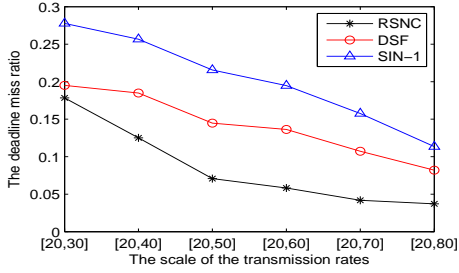}\vspace{-0.1in}
\caption{The impact of the transmission rate on the performance of the whole transmission process.}\vspace{-0.15in} \label{sim.rate}
\end{center}\vspace{-0.05in}
\end{figure}

We then investigate the impact of the transmission rate on the performance of the whole transmission. We set $n=m=10,\alpha_j=1,Tmin=10,Tmax=50$ and vary the scale of the transmission rates, i.e., $[rmin, rmax]$. As shown in Fig.~\ref{sim.rate}, the deadline miss ratio with our RSNC scheme is much lower than with other schemes. In addition, the deadline miss ratio also decreases with the increase of the transmission rate. This is because higher transmission rates incur less transmission delay, which satisfies more successful transmissions.

\vspace{-0.12in}
\subsection{The Impact of the Number of Destinations $m$}
\begin{figure}[t]
\begin{center}
\includegraphics[height=31mm,width=88mm]{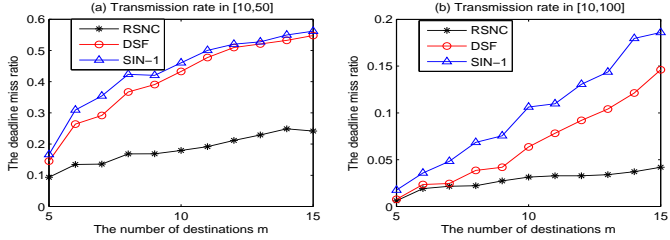}\vspace{-0.08in}
\caption{The miss deadline ratio vs. the number of destinations.}\vspace{-0.15in} \label{sim.m}
\end{center}\vspace{-0.08in}
\end{figure}

We then investigate the impact of the number of destinations $m$ and the transmission rates on the deadline miss ratio. In this case, we set $n=10,Tmin=10,Tmax=50$ by varying $m$ in $[5,15]$ for $rmin=10,rmax=50$ and $rmin=50,rmax=100$.

As shown in Fig.~\ref{sim.m}, the deadline miss ratio with our RSNC scheme is much lower than with other schemes. With the increase of $m$, the gain of our RSNC scheme increases. We can also see that the DSF algorithm does not show significant gain over SIN-1 algorithm. This is because, although with network coding in DSF, more packets can be combined together, the encoded packet may still miss its deadline at some destinations, due to inappropriate transmission rate used. From Fig.~\ref{sim.m}, we see that, with the increase of $m$, the deadline miss ratio increases. The reason is that there are more packets to be sent at $s$ within the same deadline scale.
\vspace{-0.12in}
\subsection{The Impact of the Number of Packets $n$}\vspace{-0.02in}
\begin{figure}[t]
\begin{center}\vspace{-0.08in}
\includegraphics[height=31mm,width=85mm]{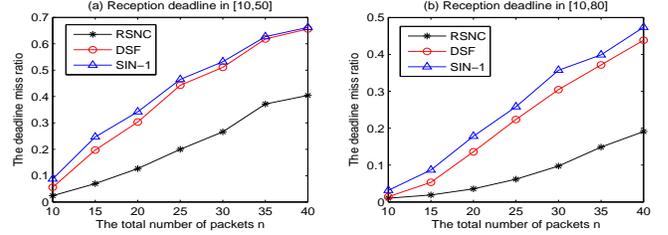}\vspace{-0.1in}
\caption{The miss deadline ratio vs. the total number of packets in P.} \vspace{-0.1in}\vspace{-0.15in}\label{sim.n}
\end{center}\vspace{-0.05in}
\end{figure}

Finally, we investigate the impact of the total number of packets $n$ and the reception deadlines on the deadline miss ratio. We set $m=10, rmin=10,rmax=50$ by varying $n$ in $[10,40]$ for the cases of $Tmin=10, Tmax=50$ and $Tmin=10,Tmax=80$.

Again, from Fig.~\ref{sim.n}, we can see that our proposed RSNC scheme has the lowest deadline miss ratio.
In addition, with the increase of $n$, the deadline miss ratio increases. This is because more packets need to be transmitted
at node $s$. From Fig.~\ref{sim.n}, it is easy to see that the deadline miss ratio is smaller when $Tmax=80$, compared with $Tmax=50$. It is reasonable because with the increase of the deadlines, less packet will lose its deadline.

\vspace{-0.12in}
\section{Conclusion}\label{Sec.conclusion}
In this paper, we propose a novel joint rate selection and network coding (RSNC) scheme for time critical applications. We first prove that the proposed problem is NP-hard, and design a novel graph model to model the problem. Using the graph model, we mathematically formulate the problem. We also propose a metric, based on which we design an efficient algorithm to determine transmission rate and coding strategy. Finally, simulation results demonstrate the proposed RSNC algorithm effectively reduces the packet deadline miss ratio.

\vspace{-0.07in}


\end{document}